\documentclass[12pt,]{article}
      \usepackage{graphicx}
      \usepackage[cp1251]{inputenc} 
\begin{document}

 \noindent {\footnotesize\it Astronomy Letters, 2010, Vol. 36, No. 11, pp. 816–822.}

 \noindent
 \begin{tabular}{llllllllllllllllllllllllllllllllllllllllllllll}
 & & & & & & & & & & & & & & & & & & & & & & & & & & & & & & & & & & & & &  \\\hline\hline
 \end{tabular}

 \vskip 0.5cm
 \centerline {\large\bf Stars Outside the Hipparcos List Closely}
 \centerline {\large\bf Encountering the Solar System}
 \bigskip
 \centerline {V.V. Bobylev}
 \bigskip
 \centerline {\small\it
 Pulkovo Astronomical Observatory, Russian Academy of Sciences,
 St-Petersburg}
 \bigskip

{\bf Abstract--}Based on currently available kinematic data, we
have searched for stars outside the Hipparcos list that either
closely encountered in the past or will encounter in the future
the Solar system within several parsecs. For the first time, we
have identified two single stars, GJ 3379 (G 099--049) and GJ 3323
(LHS 1723), as candidate for a close encounter with the solar
orbit. The star GJ 3379 could encounter the Sun more closely to a
minimum distance $d_{min} = 1.32\pm0.03$ pc at time $t_{min} =
-163\pm3$ thousand years. We have found two potential candidates
for a close encounter that have only photometrical distances: the
white dwarf SSSPM J1549--3544 without any data on its radial
velocity and the L-dwarf SDSS J1416+1348. The probabilities of
their penetration into the Oort cloud region are 0.09 (at a model
radial velocity $|V_r| = 50$ km s$^{-1}$) and 0.05, respectively.


\section{INTRODUCTION}

The Oort comet cloud (Oort 1950) is presumed to be located at the
outer boundaries of the Solar system. It is highly likely that the
cloud has a spherical shape and a radius of about $1\times10^5$ AU
(0.485 pc). Close encounters of Galactic field stars with the
Solar system play an important role in the dynamical evolution of
the Oort cloud. In particular, such passages of stars can provoke
the formation of comet showers that reach the region of the giant
planets (Emel’yanenko et al. 2007; Leto et al. 2008; Rickman et
al. 2008). Several researchers associate the traces of comet
bombardments of the Earth with the impact of such showers
(Wickramasinghe and Napier 2008).

The question about the encounters of stars with the Sun within
distances $r < 2-5$ pc was considered by Matthews (1997) and
M\"{u}ll\"{a}ri and Orlov (1996) using various ground-based
observations and by Garcia-Sanchez et al. (1999; 2001),
Dybczy\'nski (2006) and Bobylev (2010) based on Hipparcos (1997)
data in combination with stellar radial velocities. About 160
Hipparcos stars are known from the solar neighborhood 50 pc in
radius that either encountered or will encounter the Solar system
within $r<5$ pc in a time interval of $\pm$10 Myr.

The goal of this study is to search for candidate stars closely
encountering the Sun based on currently available kinematic data
on stars that do not belong to the Hipparcos catalogue. Indeed,
since the paper by M\"{u}ll\"{a}ri and Orlov (1996), who analyzed
all stars from the catalog by Gliese and Jahrei\ss~(1991), new
observational data have appeared. In this paper, we solve the
problem of statistical simulations by taking into account the
random errors in the input data in order to estimate the
probability of a star penetrating into the Oort cloud region.

\section{THE METHODS}
\subsection{Orbit construction}

We use a rectangular Galactic coordinate system with the axes
directed away from the observer toward the Galaxy center
$(l$=$0^\circ$, $b$=$0^\circ,$ the $X$ axis), in the direction of
Galactic rotation $(l=90^\circ, b=^\circ,$ the $Y$ axis), and
toward the North Pole $(b = 90^\circ$, the $Z$ axis). The
corresponding space velocity components of an object $U,V,W$ are
also directed along the $X,Y,Z$ axes. The epicyclic approximation
(Lindblad, 1927) allows the stellar orbits to be constructed in a
coordinate system rotating around the Galactic center. The
equations are (Fuchs et al., 2006)
$$ \displaylines{\hfill
 X(t)= X(0)+{U(0)\over \kappa} \sin(\kappa t)     
      +{V(0)\over 2B} (1-\cos(\kappa t)),         \hfill\llap(1)\cr\hfill
 Y(t)= Y(0)+2A \biggl( X(0)+{V(0)\over 2B}\biggr) t  
       -{\Omega_0\over B\kappa} V(0) \sin(\kappa t)
       +{2\Omega_0\over \kappa^2} U(0) (1-\cos(\kappa t)),\hfill\cr\hfill
 Z(t)= {W(0)\over \nu} \sin(\nu t) + Z(0) \cos(\nu t), \hfill
 }
$$
where $t$ is the time in Myr (we proceed from the ratio pc/Myr =
0.978 km s$^{-1}$);
 $A$ and $B$ are the Oort constants;
 $\kappa=\sqrt{-4\Omega_0 B}$ is the epicyclic frequency;
 $\Omega_0$ is the angular velocity of Galactic rotation for the local standard of
rest,
 $\Omega_0=A-B$; $\nu=\sqrt{4\pi G \rho_0}$ is the vertical oscillation
frequency, where $G$ is the gravitational constant and $\rho_0$ is
the star density in the solar neighborhood.

The parameters $X(0),Y(0),Z(0)$ and $U(0),V(0),W(0)$ in the system
of equations (1) denote the current stellar positions and
velocities. The displacement of the Sun from the Galactic plane is
taken to be $Z(0) = 17$ pc (Joshi, 2007). We calculate the
velocities $U,V,W$ relative to the local standard of rest (LSR)
with $(U,V,W)_{LSR} = (10.00, 5.25, 7.17)$ km s$^{-1}$ (Dehnen and
Binney, 1998).

At present, the question about the specific values of the Sun’s
peculiar velocity relative to the local standard of rest
 $(U,V,W)_{LSR}$ is being actively debated (Francis and Anderson, 2009;
Binney, 2010). Arguments for increasing the velocity $V_{LSR}$
from 5 km s$^{-1}$ to $\approx11$ km s$^{-1}$ are adduced. Since,
in our case, both the solar orbit and the stellar orbit are
constructed with these values, the influence of the adopted
$V_{LSR}$ is virtually imperceptible. This is confirmed by good
agreement between the encounter parameters of, for example, the
star GJ 710 and other stars obtained with various values of
$(U,V,W)_{LSR}$ by Bobylev (2010) and Garcia-Sanchez et
al.~(2001).

We took $\rho_0=0.1~M_\odot/$pc$^3$ (Holmberg and Flinn, 2004),
which gives $\nu=74$ km s$^{-1}$ kpc$^{-1}$. We used the following
Oort constants found by Bobylev et al. (2008):
 $A = 15.5\pm0.3$ km s$^{-1}$ kpc$^{-1}$ and
 $B = -12.2\pm0.7$ km s$^{-1}$ kpc$^{-1}$;
$\kappa=37$ km s$^{-1}$ kpc$^{-1}$  corresponds to these values.

Note that we neglect the gravitational interaction between the
star and the Sun.

\subsection{Statistical Simulations}

In accordance with the method of Monte Carlo statistical
simulations, we compute a set of orbits for each object by taking
into account the random errors in the input data. For each star,
we compute the encounter parameter between the stellar and solar
orbits, $d=\sqrt{\Delta X^2(t)+\Delta Y^2(t)+\Delta Z^2(t)}$). We
characterize the time of the closest encounter by two numbers,
$d_{min}$ and $t_{min}$. The errors in the stellar parameters are
assumed to be distributed normally with a dispersion $\sigma$. The
errors are added to the equatorial coordinates, proper motion
components, parallax, and radial velocity of the star. We
separately consider stars with spectroscopic distance estimates
obtained with typical errors of 20–30\%. In this case, the random
errors are added to the distance during the simulations.

\section{RESULTS AND DISCUSSION}
\subsection{Stars with Trigonometric Parallaxes}

First, let us consider the solar neighborhood about 10 pc in
radius using the list of 100 nearest stars from the Chilean RECONS
site (http://www.chara.gsu.edu/RECONS/). This list reflects the
results published before January 1, 2009. The trigonometric
parallax of each star in the list was calculated as a weighted
mean of the results of several (from one to four) observations. We
are interested in ten RECONS stars that are not Hipparcos ones
(see Table 1). Other RECONS stars are ten M dwarfs without any
radial velocity estimates and the remaining stars are Hipparcos
ones.

The equatorial coordinates and proper motions of the stars in
Table 1 were taken from Salim and Gould (2003) and L\'epine and
Shara (2005), where the proper motions were calculated by
comparing the 2MASS stellar positions with those from the Palomar
Observatory Sky Survey.

We took the radial velocities from the works of various authors.
Note that, according to the original first-class measurements by
Nidever et al.~(2002), two stars, GL 406 and GL 905, exhibit a
high stability in an observing time interval of 1--2 yr. The
internal measurement error is $\pm0.1$ km s$^{-1}$; therefore,
they were suggested as candidates for standards to determine the
radial velocities. Since these stars are M type dwarfs, the
external error in $V_r$ is $\pm0.4$ km s$^{-1}$. We took the
radial velocity of the center of mass for the triple system GL 866
ABC from the plot in Delfosse et al.~(1999). Note that, on the
whole, the radial velocities for the stars listed in the table
differ insignificantly from those in the catalog by Gliese and
Jahrei\ss~(1991). Only the star GJ 473, for which the discrepancy
in $V_r$ is two orders of magnitude, constitutes an exception. We
will discuss this situation below.

For each star, we constructed its orbit relative to the Sun in the
time interval from --2 Myr to +2 Myr. Data on six stars with an
encounter parameter $d<3$ pc are presented in Table 2 and the
trajectories of these stars are shown in Fig.1.

(1) The parameters of the GL 905 encounter with the solar orbit
found here are in good agreement with the estimates by
M\"{u}ll\"{a}ri and Orlov (1996), $d_{min} = 0.95\pm0.11$ pc and
$t_{min} = 36.3\pm1.4$ thousand years; our values have
considerably smaller random errors due to the currently available
data being very accurate.

Note that the encounter parameter for Proxima Cen is $d_{min} =
0.89\pm0.02$ pc (Bobylev, 2010) or $d_{min} = 0.95\pm0.04$ pc
(Garcia-Sanchez et al., 2001), which was obtained with a slightly
different radial velocity.  As we see from Fig.1, GL 905 has a
chance to be even slightly closer to the Sun than Proxima Cen in
$\approx$37 thousand years. At present, we know only two stars
with a smaller (than that of GL 905) encounter parameter: GJ
710=HIP 89825 with $d_{min} = 0.31\pm0.17$ pc (Bobylev, 2010) and
HIP 85661 with $d_{min} = 0.94\pm0.71$ pc (Garcia-Sanchez et al.,
2001).

(2) We have identified the stars GJ 3379 (G 099--049) and GJ 3323
(LHS 1723) as candidates for a close encounter with the solar
orbit for the first time. Their trigonometric parallaxes with a
relative error $e_\pi/\pi < 1\%$ were first determined only
recently by Henry et al.~(2006). Their radial velocities obtained
by analyzing the published data from Montes et al.~(2001) are also
accurate.

(3) As our statistical simulations show, for all six stars from
Table 2, the probability of their penetration into the Oort cloud
region is essentially zero.

\subsection{Stars with Indirect Distance Estimates}

We do not set the objective to survey all stars with
spectrophotometric distance estimates. Note only two interesting
stars, SSSPM J1549--3544 and SDSS J1416+1348, which we have been
able to reveal using data from Scholz et al.~(2004) and Schmidt et
al.~(2010a; 2010b).

(1) According to Scholz et al.~(2004), the kinematic data for
SSSPM J1549--3544 are:
 $\alpha=15^h 48^m 40^s.23$,
 $\delta=-35^\circ 44' 25''.4$,
 $\mu_\alpha \cos\delta=-591\pm8$ mas yr$^{-1}$,
 $\mu_\delta=-538\pm5$~mas yr$^{-1}$, and
 $d_{spec} = 3–4$ pc with an error of 1 pc. This star
may be the single cool white dwarf closest to the Sun (closer than
the well-known Van Maanen star, $d = 4.3$ pc). Scholz et
al.~(2004) obtained the distance $d_{spec} = 4 \pm 1$ pc using
2MASS photometry and $d_{spec} = 3\pm 1$ pc using less accurate
SSS (SuperCOSMOS Sky Survey) photometry. Since there are no radial
velocity data, the space velocities $(U,V,W)$ of this star were
estimated by Scholz et al.~(2004) for three model radial
velocities, --50, 0, and 50 km s$^{-1}$.

Following Scholz et al.~(2004), we will use $d = 4\pm1$ pc ($e_d/d
= 25\%$) and various model radial velocities for this star.
Several trajectories are shown in Fig. 2. As we see from the
figure, the sign of the radial velocity determines whether the
star could encounter the Sun in the past (positive sign) or in the
future (negative sign). We clearly see that the encounter with the
Sun becomes increasingly close with increasing magnitude of the
radial velocity. Because of the distance estimation error $\pm$1
pc, formally there is a chance of very close encounters. For
example, when the plot is shifted vertically downward by
$\approx$1 pc, the stellar trajectories highlighted by the heavy
lines fall at the boundary of the Oort cloud.

Table 3 gives two results obtained at fixed radial velocities of
the star with the addition of random errors to the proper motion
components and the distance $d$. We constructed 10 000 model
orbits for each case and in $\approx$900 cases the star falls into
the Oort cloud region, $d_{min}\leq 0.485$ pc, then $P_1 =
900/10000$.

(2) According to Schmidt et al.~(2010a, 2010b), the kinematic data
for SDSS J1416+1348 are:
 $\alpha=14^h 16^m 24^s.08$,
 $\delta=13^\circ 48' 26''.7$,
 $\mu_\alpha \cos\delta=88.0\pm2.8$ mas yr$^{-1}$,
 $\mu_\delta=139.9\pm1.3$~mas yr$^{-1}$, and
 $V_r =-42.2\pm 5.1$ km s$^{-1}$. The most reliable distance estimate, $d =
 8.0\pm1.6$ pc ($e_d/d = 20\%$), was derived by Schmidt et al.~(2010a) as
a mean of five photometric and spectroscopic determinations from
2MASS infrared data and SDSS optical data.

(3) Our statistical simulations show that both SSSPM J1549--3544
and SDSS J1416+1348 have a nonzero probability of penetrating into
the Oort cloud region: $P_1 = 0.09$ and $P_1 = 0.05,$ respectively
(the last column of Table 3). Both stars are of great interest in
the problem being solved.

We may conclude that determining the trigonometric parallaxes and
radial velocities of these stars is topical.

(4) Let us now turn to the list of 25 stars from the catalog by
Gliese and Jahrei\ss~(1991) that were revealed by M\"{u}ll\"{a}ri
and Orlov (1996) as candidates for a close encounter with the
Solar system. Since 16 of them are Hipparcos stars, the results of
the analysis of their trajectories are presented in Garcia-Sanchez
et al.~(2001) and Bobylev (2010).

Three stars are located within $d < 10$ pc; these are GJ 905, GJ
473, and GJ 3166. Since the remaining six stars are farther than
10 pc, they were not included in the RECONS list. We have already
discussed the encounter parameters of GJ 905 in the previous
section.

The star GJ 3166 (designated as No 456 NN in M\"{u}ll\"{a}ri and
Orlov, 1996) is of great interest, because, according to the
estimate by M\"{u}ll\"{a}ri and Orlov (1996), it can encounter the
Sun to a record distance $d_{min} = 0.16$ pc at time $t_{min} =
1600$ thousand years. How ever, as it turned, there is no
information about its proper motion in the catalog by Gliese and
Jahrei\ss (1991). Therefore, its space velocities $U,V,W$ were
calculated by assuming the tangential velocity to be zero,
$V_t=0.$

Taking $d = 20.8$ pc (photometric distance), $\mu_t = 0$ mas
yr$^{-1}$, and $V_r = -12$ km s$^{-1}$ for GJ 3166, as in the
catalog by Gliese and Jahrei\ss (1991), we find the encounter
parameters $d_{min} = 0.18$ pc and $t_{min} = 1708$ thousand
years, which confirm the result by M\"{u}ll\"{a}ri and Orlov. For
the same distance and radial velocity but taking the proper motion
components
 $\mu_\alpha\cos\delta = -103.3 \pm 8.1$ mas yr$^{-1}$ and
 $\mu_\delta = -67.7 \pm 9.2$ mas yr$^{-1}$ from the UCAC3 catalog
(Zacharias et al., 2009), we find completely different encounter
parameters, $d_{min} = 15$ pc and $t_{min} = 590$ thousand years,
which are already less interesting in our problem. Note that the
absolute proper motions of this star are also available in the XPM
catalog (Fedorov et al., 2009):
 $\mu_\alpha\cos\delta=-87.2$ mas yr$^{-1}$ and
 $\mu_\delta=-68.1$ mas yr$^{-1}$.
These were determined by comparing the 2MASS and Palomar
Observatory Sky Survey positions referenced to galaxies with a
mean error of about 6 mas yr$^{-1}$ in each coordinate (Bobylev et
al., 2010).

Finally, according to the estimate by M\"{u}ll\"{a}ri and Orlov
(1996), GJ 473 can encounter the Sun very closely, $d_{min} =
0.29$ pc, at time $t_{min} = 7.5$ thousand years. These parameters
were obtained using the radial velocity $V_r = -553.7$ km s$^{-1}$
(Gliese and Jahrei\ss, 1991). GJ 473 (LHS 333=FL Vir=Wolf 424 AB)
is a close binary system with a known orbit (Torres et al., 1999).
The measurements by Tinney and Reid (1998) performed with a
high-resolution spectrometer yield the system’s heliocentric
radial velocity $V_r = 0.9 \pm 1.7$ km s$^{-1}$. Note that
previous measurements also gave a low radial velocity for this
system: $V_r = -5 \pm 5$ km s$^{-1}$ (GCRV, Wilson 1953). The
radial velocity in the compilation by Gliese and Jahrei\ss (1991)
is probably erroneous. Since the encounter parameters of GJ 473
calculated with its new radial velocity $V_r = 0.9 \pm 1.7$ km
s$^{-1}$, $d_{min} = 6 \pm 5$ pc and $t_{min} = -3 \pm 6$ thousand
years, are no longer the close encounter parameters, this star was
not included in Table 2.

\section{CONCLUSIONS}

Based on currently available kinematic data, we searched for stars
that either encountered or will encounter the solar neighborhood
within less than 3 pc. We considered stars outside the Hipparcos
list. For each of them, there is an estimate of its trigonometric
parallax with a relative error $e_\pi/\pi < 2\%,$ radial velocity,
and proper motion components. We found six such stars that are an
important supplement to the list of Hipparcos stars closely
encountering the Solar system (Garcia-Sanchez et al., 2001;
Bobylev, 2001).

For the first time, two single stars, GJ 3379 (G 099--049) and GJ
3323 (LHS 1723), have been identified as candidates for a close
encounter with the solar orbit.

We confirmed the remarkable result by M\"{u}ll\"{a}ri and Orlov
(1996) that the star GL 905 could encounter the Sun fairly
closely: $d_{min} = 0.93\pm 0.01$ pc and $t_{min} = 37.1\pm0.2$
thousand years.

Two unique stars are located in the immediate solar neighborhood
($d < 10$ pc)---the white dwarf SSSPM J1549--3544 and the L dwarf
SDSS J1416+1348. Our statistical simulations showed that both of
them have a nonzero probability of penetrating into the Oort cloud
region: $P_1 = 0.09$ and $P_1 = 0.05$, respectively. Determining
the trigonometric parallaxes and radial velocities for these stars
is topical. This task can be accomplished using both ground-based
and space (e.g., GAIA) observations.

Based on new data, we showed that the stars GJ 473 and GJ 3166 are
not suitable candidates for a very close encounter with the Solar
system as was presumed previously.

\bigskip
{\bf ACKNOWLEDGMENTS}

 I am grateful to A.T. Bajkova for the software package for
statistical simulations. The SIMBAD database and the RECONS site
were very helpful in the work. This work was supported by the
Russian Foundation for Basic Research (project nos. 08--02--00400
and 09--02--90443--Ukr\_f) and in part by the ``Origin and
Evolution of Stars and Galaxies'' Program of the Presidium of the
Russian Academy of Sciences.

\bigskip
{\bf REFERENCES}

 {\small

  M. Barbier-Brossat and P. Figon, Astron. Astrophys. Suppl. Ser. 142, 217 (2000).

J.J. Binney, Mon. Not. R. Astron. Soc. 401, 2318 (2010).

V.V. Bobylev, Pis'ma Astron. Zh. 36, 230 (2010) [Astron. Lett. 36,
220 (2010)].

V.V. Bobylev, A.T. Bajkova, and A.S. Stepanishchev, Pis'ma Astron.
Zh. 34, 570 (2008) [Astron. Lett. 34, 515 (2008)].

V.V. Bobylev, P.N. Fedorov, A.T. Bajkova, and V.S. Akhmetov,
Pis'ma Astron. Zh.36, 535 (2010) [Astron. Lett. 36, 506 (2010)].

W. Dehnen and J.J. Binney, Mon. Not. R. Astron. Soc. 298, 387
(1998).

X. Delfosse, T. Forveille, S. Udry, et al., Astron. Astrophys.
350, L39 (1999).

P.A. Dybczy\'nski, Astron. Astrophys.  449, 1233 (2006).

P.N. Fedorov, A.A. Myznikov and V.S. Akhmetov, Mon. Not. R.
Astron. Soc. 393, 133 (2009).

C. Francis and E. Anderson, New Astron. 14, 615 (2009).

B. Fuchs, D. Breitschwerdt, M.A. Avilez, et al., Mon. Not.R .
Astron. Soc. 373, 993 (2006).

J. Garcia-S\'anchez, R.A. Preston, D.L. Jones, et al., Astron.J.
117, 1042 (1999).

J. Garcia-S\'anchez, P.R. Weissman, R.A. Preston, et al., Astron.
Astrophys. 379, 634 (2001).

W. Gliese and H. Jahrei\ss, Preliminary Version of the Third
Catalogue of Nearby Stars (CNS3), Computer-Readable Version on ADC
Selected Astronomical Catalogs, Vol.1, CD-ROM (1991).

T.J. Henry, W.-C. Jao, J.P. Subasavage, et al., Astron. J. 132,
2360 (2006).

J. Holmberg and C. Flinn, Mon. Not. R. Astron. Soc. 352, 440
(2004).

Y.C. Joshi, Mon. Not. R. Astron. Soc. 378, 768 (2007).

S. L\'epine and M.M. Shara, Astron. J. 121, 1483 (2005).

G. Leto, M. Jakubik, T. Paulech, et al., Mon. Not. R. Astron. Soc.
391, 1350 (2008).

B. Lindblad, Arkiv Mat., Astron., Fysik A 20, No 17 (1927).

R.A.J. Matthews, R. Astron. Soc. Quart. J. 35, 1 (1994).

D. Montes, J. L\'opez-Santiago, M.C. G\'alvez, et al., Mon. Not.
R. Astron. Soc. 328, 45 (2001).

A.A. M\"{u}ll\"{a}ri and V.V. Orlov, Earth, Moon, and Planets
(Kluwer, Netherlands, 1996), v.72, p. 19.

D.L. Nidever, G.W. Marcy, R.P. Butler, et al., Astrophys. J.
Suppl. Ser. 141, 503 (2002).

J.H. Oort, Bull. Astron. Inst. Netherland 11, 91 (1950).

H. Rickman, M. Fouchard, C. Froeschl\'e , et al., Sel. Mech. Dyn.
Astr. 102, 111 (2008).

S. Salim and A. Gould, Astrophys. J. 582, 1011 (2003).

S. Schmidt, A.A. West, A. Burgasser, et al., astroph/ 0912.3565
(2010a).

S. Schmidt, A.A. West, S. Hawley , et al., astroph/ 1001.3402
(2010b).

R.-D. Scholz, I. Lehmann, I. Matute, et al., Astron. Astrophys.
425, 519 (2004).

The HIPPARCOS and Tycho Catalogues, ESA SP-1200 (1997).

C.G. Tinney and I.N. Reid, Mon. Not. R. Astron. Soc. 301, 1031
(1998).

G. Torres, T. Henry, O. Franz, et al., Astron. J. 117, 562 (1999).

J.T. Wickramasinghe and W.M. Napier, Mon. Not. R. Astron. Soc.
387, 153 (2008).

R.E. Wilson, General Catalogue of Stellar Velocity, Publ. 601
(Carnegie Inst., Washington, DC, 1953).

N. Zacharias, C. Finch, T. Girard, et al., CDS Strasbourg, I/315
(2009).

\bigskip
Translated by Shtaerman

}

\newpage
{
\begin{table}[t]                                                
\caption[]{\small\baselineskip=1.0ex\protect
 Stars and multiple systems with trigonometric parallaxes
}
\begin{center}
\begin{tabular}{|l|r|r|r|r|c|}\hline

           &                      &                             &             &           &       \\
      Star &    $\alpha_{J2000},$ & $\mu_\alpha,$ mas yr$^{-1}$ & $\pi,$ mas  & $V_r,$ km s$^{-1}$ & Ref*  \\
           &    $\delta_{J2000}$  & $\mu_\delta,$ mas yr$^{-1}$ &             &           &       \\
&&&&&\\\hline

 GJ 65 AB  & $  1^h 39^m 01^s.54$ & $ 3296\pm5$ & $373.7\pm2.7$ & $29.0\pm4.6$  & e \\
           & $-17^o 57'  01''.8$  & $  563\pm5$ &&&\\\hline
 GJ 299    & $  8^h 11^m 57^s.58$ & $ 1099\pm8$ & $146.3\pm3.1$ &  $-35\pm5$    & e \\
           & $  8^o 46'  22''.1$  & $-5123\pm8$ &&&  \\\hline
 GJ 388 AB & $ 10^h 19^m 36^s.28$ & $ -502\pm8$ & $204.6\pm2.8$ & $11.6\pm0.3$  & e \\
           & $ 19^o 52'  12''.1$  & $  -43\pm8$ &&&\\\hline
 GJ 406    & $ 10^h 56^m 28^s.86$ & $-3842\pm8$ & $419.1\pm2.1$ & $19.5\pm0.4$  & a \\
           & $  7^o 00'  52''.8$  & $-2727\pm8$ &&&\\\hline
 GJ 473 AB & $ 12^h 33^m 17^s.38$ & $-1730\pm8$ & $227.9\pm4.6$ &  $0.9\pm1.7$  & d \\
           & $  8^o 46'  22''.1$  & $  203\pm8$ &&&  \\\hline
GJ 866 ABC & $ 22^h 38^m 33^s.73$ & $ 2314\pm5$ & $289.5\pm4.4$ &   $-50\pm1$   & c \\
           & $-15^o 17'  57''.3$  & $ 2295\pm5$ &&&\\\hline
 GJ 905    & $ 23^h 41^m 54^s.99$ & $  111\pm8$ & $316.0\pm1.1$ & $-78.0\pm0.4$ & a \\
           & $ 44^o 10'  40''.8$  & $-1584\pm8$ &&&  \\\hline
 GJ 1111   & $  8^h 29^m 49^s.35$ & $-1112\pm5$ & $275.8\pm3.0$ &   $9.0\pm0.5$ & b \\
           & $ 26^o 46'  33''.7$  & $ -611\pm4$ &&&\\\hline
 GJ 3323   & $  5^h 01^m 57^s.47$ & $ -550\pm5$ & $187.9\pm1.3$ &  $42.0\pm0.1$ & b \\
           & $ -6^o 56'  45''.9$  & $ -533\pm5$ &&&\\\hline
 GJ 3379   & $  6^h 00^m 03^s.50$ & $  311\pm3$ & $190.9\pm1.9$ &  $30.0\pm0.1$ & b \\
           & $  2^o 42'  23''.67$ & $  -42\pm3$ &&&\\\hline

\end{tabular}
\end{center}
 {\small
Note. The radial velocities were taken from the following papers:
(a) Nidever (2002), (b) Montes et al.( 2001), (c) Delfosse et al.
(1999), (d) Tinney and Reid (1998), (e) Barbier-Brossat and Figon
(2000); the parallaxes are given according to the RECONS list of
100 nearest stars.

 }\vskip4mm
\end{table}
 }

{
\begin{table}[t]                                                
\caption[]{\small\baselineskip=1.0ex\protect
 Candidates for a close encounter with the Solar system
}
\begin{center}
\begin{tabular}{|l|c|c|c|c|}\hline

            &             &             &            &             \\
       Star & SP        & $M/M_\odot$ & $d_{min},$ pc & $t_{min},$ thousand yr  \\
&&&&\\\hline

 GJ 905     & M5.5V       & 0.12           & $0.93\pm0.01$ & $ 37.1\pm0.2$ \\\hline
 GJ 3379    & M3.5V       & 0.19           & $1.32\pm0.03$ & $ -163\pm3$ \\\hline
 GJ 65 AB   & M5.5V/M6.0V & 0.11/0.10      & $2.21\pm0.12$ & $  -29\pm2$ \\\hline
 GJ 406     & M6.0V       & 0.09           & $2.24\pm0.02$ & $  -15\pm1$ \\\hline
 GJ 3323    & M4.5V       & 0.15           & $2.25\pm0.05$ & $ -104\pm3$ \\\hline
 GJ 866 ABC & M5.0V/--/-- & 0.11/0.11/0.10 & $2.52\pm0.07$ & $ 32.3\pm0.3$ \\\hline

\end{tabular}
\end{center}
 {\small
 }\vskip4mm
\end{table}
 }

{
\begin{table}[t]                                                
\caption[]{\small\baselineskip=1.0ex\protect
 Stars with spectroscopic or photometric distance
estimates
 }
\begin{center}
\begin{tabular}{|c|c|c|c|c|c|c|c|c|}\hline

         &       &             &            &            &        &            \\
    Star &   SP  & $M/M_\odot$ & $d_{min},$ & $t_{min},$ & $V_r,$ & $P_1$ \\
         &       &             &     pc     & thousand yr   &   km s$^{-1}$ &            \\\hline

 SSSPM~J1549$-$3544 & $>$DC11 & $\approx$ 0.5  & $1.21\pm0.58$ & $-72\pm6$  & $+50$  & 0.09 \\\hline
 SSSPM~J1549$-$3544 &        &                & $1.21\pm0.58$ & $+72\pm6$  & $-50$  & 0.09 \\\hline
 SDSS~J1416+1348    &   L5V  & $\approx$ 0.08 & $1.24\pm0.65$ & $186\pm44$ & $-42.2\pm5.1$& 0.05 \\\hline

\end{tabular}
\end{center}
 {\small
 }\vskip4mm
\end{table}
 }

\newpage
\begin{figure}[t]{
\begin{center}
  \includegraphics[width=160mm]{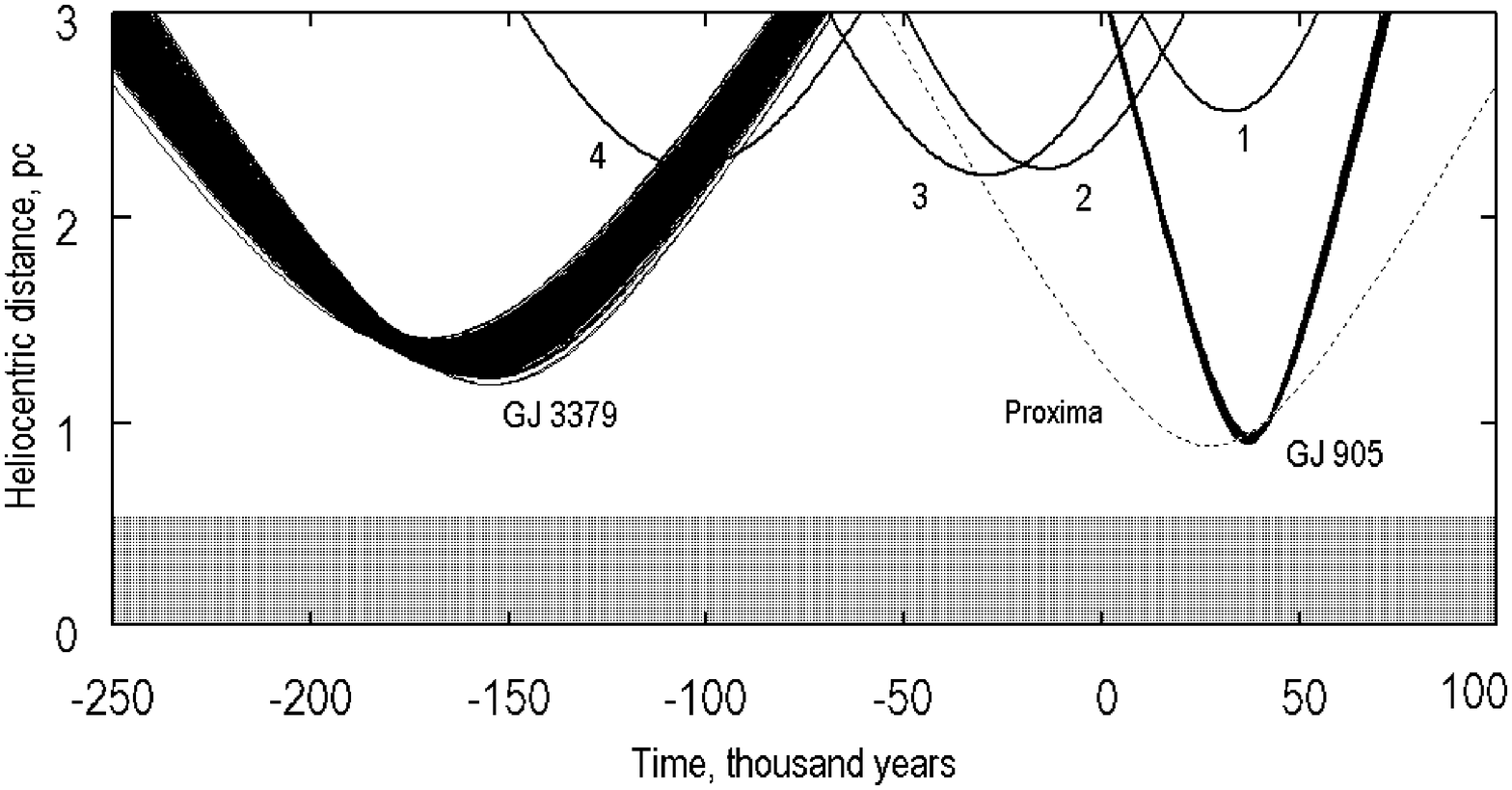}
\vskip0.3mm
\end{center}}
\end{figure}
Fig. 1. Trajectories of the stars relative to the Sun: (1) GJ 866,
(2) GJ 406, (3) GJ 65, (4) GJ 3323; for GJ 905 and GJ 3379, we
give the model trajectories computed by taking into account the
random errors in the observational data (1000 realizations). The
trajectories hatch the 3$\sigma$ confidence regions, the Oort
cloud region is shaded, the dotted line indicates the trajectory
of Proxima Cen.

\newpage
\begin{figure}[t]{
\begin{center}
  \includegraphics[width=160mm]{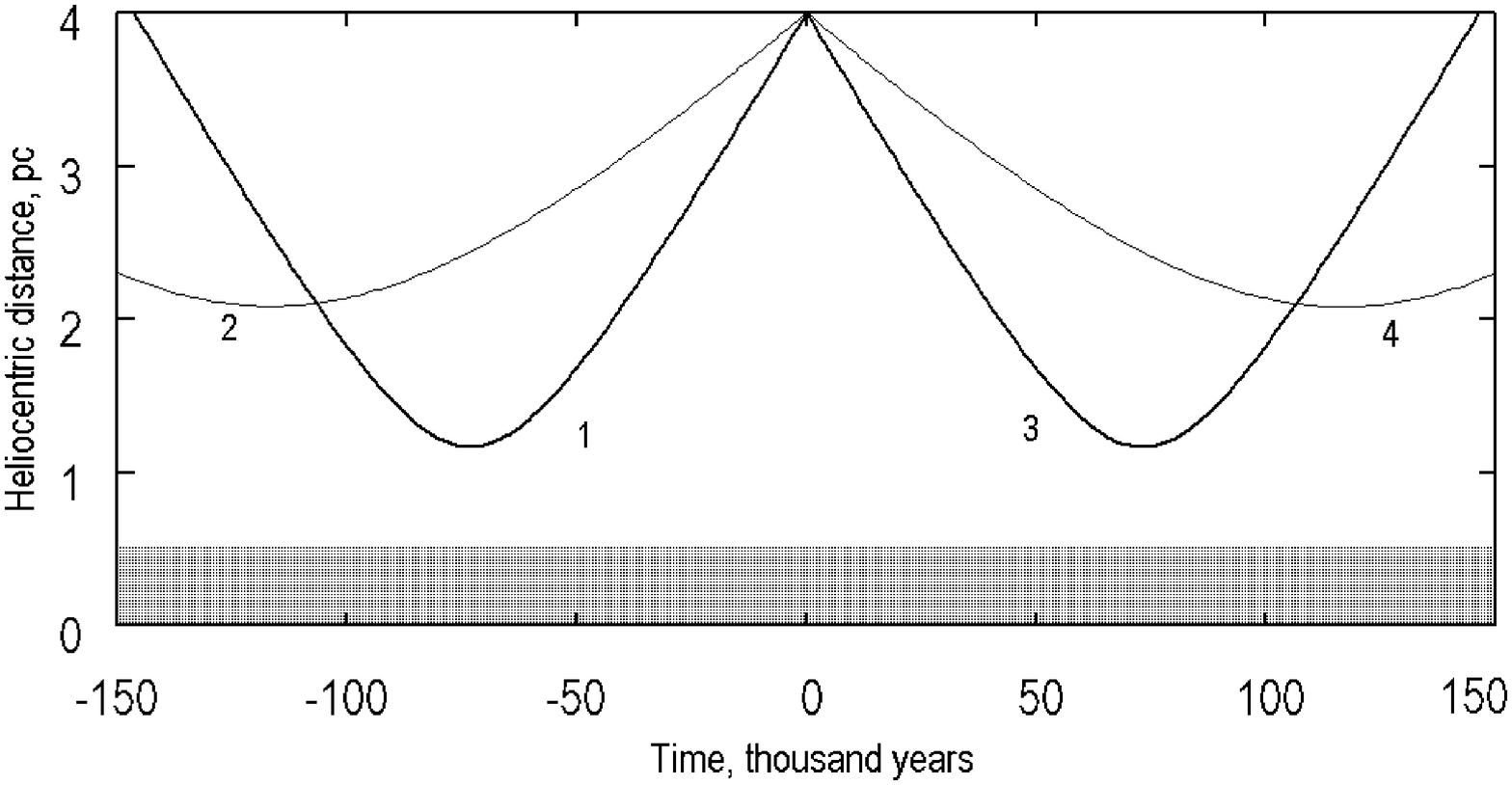}
\vskip0.3mm
\end{center}}
\end{figure}

Fig. 2. Model trajectories of the star SSSPM J1549–3544 relative
to the Sun: $d = 4$ pc at $V_r = +50$ km s$^{-1}$ (1), $V_r = +25$
km s$^{-1}$ (2), $V_r = -50$ km s$^{-1}$ (3), and $V_r = -25$ km
s$^{-1 }$ (4).

\end{document}